\documentclass[aps,prd,twocolumn,groupedaddress,showpacs]{revtex4-1}

\usepackage{graphicx}
\usepackage{color}
\usepackage[T1]{fontenc}
\usepackage{amssymb,amsmath}
\usepackage{ulem}

\newcommand{\bfv}{\mbox{\boldmath $v$}}
\newcommand{\bfV}{\mbox{\boldmath $V$}}

\begin{document}


\title{On the relative velocity distribution for general statistics and an application to big-bang nucleosynthesis under Tsallis statistics}


\author{Motohiko Kusakabe$^{1,2}$}
\email{kusakabe@buaa.edu.cn}
\author{Toshitaka Kajino$^{1,2,3}$}
\author{Grant J. Mathews$^{2,4}$}
\author{Yudong Luo$^{2,3}$}
\affiliation{
$^1$School of Physics and Nuclear Energy Engineering, and
  International Research Center for Big-Bang Cosmology and Element Genesis, Beihang University 37, Xueyuan Rd., Haidian-qu, Beijing 100083 China}
\affiliation{
  $^2$National Astronomical Observatory of Japan, 2-21-1 Osawa, Mitaka, Tokyo 181-8588, Japan}
\affiliation{
  $^3$Graduate School of Science, University of Tokyo, 7-3-1 Hongo, Bunkyo-ku, Tokyo 113-0033, Japan}
\affiliation{
  $^4$Center for Astrophysics, Department of Physics, University of Notre Dame, Notre Dame, IN 46556, U.S.A.}


\date{\today}

\begin{abstract}
The distribution function of the relative velocity in a two-body reaction of nonrelativistic uncorrelated particles is derived for general cases of given distribution functions of single particle velocities. The distribution function is then used in calculations of thermonuclear reaction rates. As an example, we take the Tsallis non-Maxwellian distribution, and show that the distribution function of the relative velocity is different from the Tsallis distribution. We identify an inconsistency in previous studies of nuclear reaction rates within Tsallis statistics, and derive revised nuclear reaction rates. Utilizing the revised rates, accurate results of big bang nucleosynthesis are obtained for the Tsallis statistics. For this application it is more difficult to reduce the primordial $^7$Li abundance while keeping other nuclear abundances within the observational constraints. A small deviation from a Maxwell-Boltzmann distribution can increase the D abundance and slightly reduce $^7$Li abundance. Although it is impossible to realize a $^7$Li abundance at the level observed in metal-poor stars, a significant decrease is possible while maintaining a consistency with the observed D abundance. 
\end{abstract}


\maketitle


\section{Introduction}\label{sec1}

Deviations of particle distribution functions from the Maxwell-Boltzmann (MB) distribution are often found in geophysical and astrophysical observations of systems out of equilibrium \cite{Vasyliunas1968,Bame1967,Livadiotis2009}. For example, a power-law distribution has been observed in the electron spectrum of the magnetosphere \cite{Vasyliunas1968}. A power-law distribution called the kappa-distribution is also realized in Tsallis's statistical model in which a generalization of the entropy is postulated to be $S_q \equiv k (1-\sum_{i=1}^W p_i^q )/(q-1)$, where $q$ is a real parameter, $k$ is Boltzmann's constant, and $p_i$ are the probabilities of $i$ with the total configuration number $W$ \cite{Tsallis1988}. See Ref. \cite{Livadiotis2009} for formulations of the Tsallis statistics and relations between the Tsallis distribution function and the power-law. The Tsallis distribution function for $q<1$ has also been found to be realized by a special pattern of temperature fluctuation \cite{Wilk:1999dr}. Moreover, 
non-relativistic baryons in equilibrium with a thermal bath of relativistic electrons have been proposed \cite{Sasankan:2018pfq} to obey modified MB statistics. 

Usually, one   assumes  a MB distribution for nonrelativistic particles in calculations of thermonuclear reaction rates. Especially, during big bang nucleosynthesis (BBN) and stellar nuclear burning, the temperature is high enough that very frequent scatterings quickly realize the MB distribution of nuclei (e.g., \cite{Voronchev:2008zz,Voronchev:2009zz}). Therefore, the standard BBN (SBBN) theory is based upon the MB nuclear distribution. However, in nonstandard BBN models such as those involving scattering from relativistic electrons \cite{Sasankan:2018pfq}, or the injection of non-thermal photons \cite{Lindley1979MNRAS.188P..15L} and hadrons \cite{Dimopoulos:1987fz,Reno:1987qw}  the particle distribution functions can deviate significantly from a MB distribution.

Bertulani et al. \cite{Bertulani:2012sv} analyzed effects of the Tsallis distribution function 
for nuclear velocities on BBN. They assumed the Tsallis distribution for the whole energy range and found that the thermonuclear reaction rates strongly depend on the Tsallis parameter $q$. They then concluded that the Tsallis parameter should be very close to unity which corresponds to the MB distribution in order to satisfy observational constraints on light element abundances.
An extended parameter search was made with modifications to the 2-body reverse reaction rates taken into account \cite{Hou:2017uap}. It was found that a slightly softer spectra than the MB distribution results in a decrease of the $^7$Li abundance. 

In the SBBN model, theoretical primordial abundances are consistent with observational constraints except for the Li abundance \cite{Coc:2014oia,Cyburt:2015mya,Mathews:2017xht}. The discrepancy between the theoretical prediction of the primordial Li abundance and astronomical observations of metal-poor stars \cite{Spite:1982dd,Ryan:1999vr,Melendez:2004ni,Asplund:2005yt,Bonifacio:2006au,shi2007,Aoki:2009ce,Hernandez:2009gn,Sbordone:2010zi,Monaco:2010mm,Monaco:2011sd,Mucciarelli:2011ts,Aoki:2012wb,Aoki2012b} is called the Li problem.
Therefore, this non-MB distribution is a possible solution to the Li problem of metal-poor stars \cite{Coc:2014oia,Cyburt:2015mya,Mathews:2017xht}.

In this paper, we derive a new formulation of the relative velocity distribution function for general distributions for reacting nonrelativistic particles.
In Sec. \ref{sec2}, the relative velocity distribution function is formulated, and an explicit function is shown for the case of Tsallis statistics.
In Sec. \ref{sec3}, the relative velocity distribution is calculated for Tsallis statistics, and we show that our relative velocity distribution is different from that adopted in previous studies. In addition, thermonuclear reaction rates and primordial light element abundances are calculated under the Tsallis statistics.
In Sec. \ref{sec4}, we summarize this study.

We note, however, that the correction derived here  applies to a variety of physical mechanisms that can directly modify the particle velocity distribution functions. There are many examples in the literature of physical mechanisms that lead to such altered velocity distribution functions.  For example, such Tsallis distribution functions have been shown to arise from the combined effects of local magnetic and electric fields \cite{Rossani2009}, or in space plasmas \cite{Pavlos2012} including the  heliosheath \cite{Burlaga2006}, planetary magnetospheres,  the solar corona, solar dynamics,  and cosmic rays.  We note, however, that a modified Tsallis-like distribution may also approximate the effect of relativistic electron scattering during BBN \cite{Sasankan:2018pfq} or sub-horizon temperature fluctuations as in \cite{Luo2018}.  In this latter case the sub-horizon local velocity distributions  are indeed Maxwell-Boltzmann and a Tsallis-like distribution only emerges when averaging over a set of sub-horizon volumes at different temperature.   In such models, a direct application of the Tsallis distribution to the relative velocity distribution as in Ref.~\cite{Bertulani:2012sv}  is justified.
 
\section{Model}\label{sec2}

Our model for the relative velocity distribution for general statistics assumes three things: 1) Each nucleus is described by a non-Maxwell-Boltzmann velocity distribution; 2)  conservation of momentum;  and 3) conservation of energy.  We show in this section that imposing these conditions leads to a general relative velocity distribution function that does not resemble the original distribution  in which velocities are replaced with the relative center-of-mass velocities and masses are replaced by the reduced mass.   To begin with the thermal rate for a two-body reaction of nonrelativistic uncorrelated particles is given by
\begin{equation}
  \langle \sigma v \rangle =\int d \bfv_1 f(\bfv_1) \int d \bfv_2 f(\bfv_2)
  \sigma(E) v,
  \label{eq1}
\end{equation}
where
$\sigma$ is the reaction cross section,
$\bfv_i$ is the velocity vector of species $i=1$ and 2,
$f(\bfv_i)$ is the velocity distribution function of $i$,
$v=|\bfv_1 -\bfv_2|$ is the relative velocity, and
$E =\mu v^2 /2$ is the center of mass (CM) energy.

We use the CM parameter transformations as follows:
\begin{eqnarray}
  M &=& m_1 +m_2 \\
  \label{eq2}
  \mu &=& \frac{m_1 m_2}{m_1 +m_2} \\
  \label{eq3}
  \bfv &=& \bfv_1 -\bfv_2 \\
  \label{eq4}
  \bfV &=&\frac{ m_1 \bfv_1 +m_2 \bfv_2}{m_1 +m_2}.
  \label{eq5}
\end{eqnarray}
There is also a conservation of energy relation,
\begin{eqnarray}
  m_1 \bfv_1^2 +m_2 \bfv_2^2
  = M V^2 +\mu v^2.
  \label{eq6}
\end{eqnarray}

We use the variable transformations from $\bfv_1$ and $\bfv_2$ to $\bfv$ and $\bfV$ (e.g., \cite{Clayton1984}). We define the distribution function of the relative velocity $\bfv$ that satisfies
\begin{eqnarray}
  \int d \bfv_1 f(\bfv_1) \int d \bfv_2 f(\bfv_2)
  &=&\int d \bfv d \bfV  f(\bfv_1) f(\bfv_2) \nonumber \\
  &=&\int d \bfv f^{\rm rel}(\bfv).
  \label{eq7}
\end{eqnarray}
The distribution function $f^{\rm rel}(\bfv)$ is then given by
\begin{equation}
  f^{\rm rel}(\bfv) =\int d \bfV  \left[ f(\bfv_1) f(\bfv_2) \right]_{\bfv},
  \label{eq8}
\end{equation}
where the quantity in brackets with the subscript $\bfv$ is estimated for a fixed $\bfv$.

\subsection{the Maxwell-Boltzmann distribution}\label{sec2a}
The MB distribution is given \cite{Tsallis1988,Bertulani:2012sv,Hou:2017uap} by
\begin{equation}
  f_{\rm MB}(\bfv_i) =\left( \frac{m_i}{2 \pi k T} \right)^{3/2}
  \exp\left( -\frac{m_i v_i^2}{2k T} \right),
  \label{eq9}
\end{equation}
      where
      $T$ is the temperature.
If both of the reacting particles are described by an MB distribution, i.e., $f_{\rm MB}$, the integration over the velocities is given by
\begin{eqnarray}
  && \int d \bfv_1 f_{\rm MB}(\bfv_1) \int d \bfv_2 f_{\rm MB}(\bfv_2) \nonumber \\
  &=& \frac{\left( m_1 m_2 \right)^{3/2}}{\left( 2 \pi k T \right)^3}
  \int \exp\left[ -\frac{M V^2 +\mu v^2}{2 kT} \right]
  d \bfv d \bfV  \nonumber \\
  &=& \frac{\mu^{3/2}}{\left( 2 \pi k T \right)^{3/2}}
  \int \exp\left[ -\frac{\mu v^2}{2 kT} \right]
  d \bfv.
  \label{eq10}
\end{eqnarray}
Therefore, the CM distribution function of the relative velocity for the case of MB statistics has the same form as that of the
individual particle distribution functions, i.e.
\begin{equation}
  f_{\rm MB}^{\rm rel}(\bfv) =
  \frac{\mu^{3/2}}{\left( 2 \pi k T \right)^{3/2}}
  \exp\left[ -\frac{\mu v^2}{2 kT} \right].
  \label{eq11}
\end{equation}

\subsection{Tsallis distribution}\label{sec2b}
The Tsallis distribution is, however, given by
\begin{equation}
  f_q(\bfv_i) =B_q(m_i c^2/kT) \left( \frac{m_i}{2 \pi k T} \right)^{3/2}
  \left[1 -\left( q -1 \right) \frac{m_i v_i^2}{2k T} \right]^{1/(q-1)},
  \label{eq12}
\end{equation}
where $B_q(m_i c^2/kT)$ is a normalization constant determined from the requirement $\int f_q(\bfv_i) d \bfv_i =1$.
When both of the reacting particles are described by this distribution with the same $q$ value, the product of the two distribution functions is given by
\begin{eqnarray}
  f_q(\bfv_1) f_q(\bfv_2)
  &=& B_q(m_1 c^2/kT) B_q(m_2 c^2/kT)
  \frac{\left( m_1 m_2 \right)^{3/2}}{\left( 2 \pi k T \right)^3} \nonumber \\
  && \times
  \left[1 -\left( q -1 \right) \frac{m_1 v_1^2}{2k T} \right]^{1/(q-1)}  \nonumber \\
  && \times
  \left[1 -\left( q -1 \right) \frac{m_2 v_2^2}{2k T} \right]^{1/(q-1)}
  \label{eq13}
\end{eqnarray}
We then have the following transformation:
\begin{widetext}
\begin{eqnarray}
  I_q(v_1, v_2; m_1, m_2, T) &=&
  \left[1 -\left( q -1 \right) \frac{m_1 v_1^2}{2k T} \right]^{1/(q-1)}
  \left[1 -\left( q -1 \right) \frac{m_2 v_2^2}{2k T} \right]^{1/(q-1)} \\
  \label{eq14}
  &=& \left[1 -\left( q -1 \right) \frac{m_1 v_1^2 +m_2 v_2^2}{2k T}
    +\left( q -1 \right)^2 \frac{m_1 m_2 v_1^2 v_2^2}{(2k T)^2} \right]^{1/(q-1)} \\
  \label{eq15}
  &=& \left[1 -\left( q -1 \right) \frac{M V^2 +\mu v^2}{2k T}
    +\left( q -1 \right)^2 \frac{m_1 m_2 v_1^2 v_2^2}{(2k T)^2} \right]^{1/(q-1)},
  \label{eq16}
\end{eqnarray}
\end{widetext}
where $v_1^2$ and $v_2^2$ can be expressed as functions of $\bfV$ and $\bfv$ by
\begin{eqnarray}
  v_1^2 &=&V^2 +2 \frac{m_2}{M} \bfV \cdot \bfv +\frac{m_2^2}{M^2} v^2 \\
  \label{eq17}
  v_2^2 &=&V^2 -2 \frac{m_1}{M} \bfV \cdot \bfv +\frac{m_1^2}{M^2} v^2.
  \label{eq18}
\end{eqnarray}

The distribution function of the relative velocity for the Tsallis particles is then given by
\begin{equation}
  f_q^{\rm rel}(\bfv) =
  \int \left[ f_q(\bfv_1) f_q(\bfv_2) \right]_{\bfv} d \bfV.
  \label{eq19}
\end{equation}
The integration ranges of $V$ and $v$ are derived as follows. For $q>1$, the velocities for two species are limited \cite{Livadiotis2009} to be 
\begin{equation}
  m_i v_i^2 \leq \frac{2 kT}{q-1}.
  \label{eq20}
\end{equation}
The integration ranges are then given by
\begin{eqnarray}
  0 \leq& V & \leq \sqrt{\mathstrut \frac{2 kT}{q-1}}
  \frac{\sqrt{\mathstrut m_1} +\sqrt{\mathstrut m_2}}{M} \\
  \label{eq21}
  0 \leq& v & \leq v_{1,{\rm max}} +v_{2,{\rm max}} =\sqrt{\mathstrut \frac{2kT}{q-1}} \left( \frac{1}{\sqrt{\mathstrut m_1}} + \frac{1}{\sqrt{\mathstrut m_2}} \right). \nonumber \\
  \label{eq22}
\end{eqnarray}
For a fixed $\bfv$, the distribution function of $v$ is then transformed to
\begin{widetext}
\begin{eqnarray}
  f_q^{\rm rel}(\bfv) &=&
  2 \pi \int_{-1}^1 d \cos\theta \int_0^{V_{\rm max}} V^2 dV
  f_q(\bfv_1) f_q(\bfv_2) \nonumber \\
  &=&
  2 \pi
  B_q(m_1 c^2/kT) B_q(m_2 c^2/kT)
  \frac{\left( m_1 m_2 \right)^{3/2}}{\left( 2 \pi k T \right)^3}
  \int_{-1}^1 d \cos\theta \int_0^{V_{\rm max}} V^2 dV
  I_q(V, \cos\theta; m_1, m_2, T, v),
  \label{eq23}
\end{eqnarray}
where
the function $I_q$ is given by
\begin{eqnarray}
  I_q(V, \cos\theta; m_1, m_2, T, v) &=&
  \left\{
  \begin{array}{l}
  \left[1 -\left( q -1 \right) \frac{m_1 v_1^2}{2k T} \right]^{1/(q-1)}
  \left[1 -\left( q -1 \right) \frac{m_2 v_2^2}{2k T} \right]^{1/(q-1)} \\
  ~~~~~~~~~~(1 -\left( q -1 \right) \frac{m_1 v_1^2}{2k T} > 0~{\rm and}~
  1 -\left( q -1 \right) \frac{m_2 v_2^2}{2k T} >0) \\
  0~~~~~~~~~~({\rm otherwise}) \\
  \end{array}
  \right. \\
  \label{eq24}
  v_1^2 &=&V^2 +2 \frac{m_2}{M} V v \cos\theta +\frac{m_2^2}{M^2} v^2 \\
  \label{eq25}
  v_2^2 &=&V^2 -2 \frac{m_1}{M} V v \cos\theta +\frac{m_1^2}{M^2} v^2.
  \label{eq26}
\end{eqnarray}
\end{widetext}

This distribution function is manifestly different from that employed in previous studies \cite{Bertulani:2012sv,Hou:2017uap}, which adopted exactly the same Tsallis function as the single particle velocity distribution with the mass replaced by the reduced mass. Thus, $E_1 E_2$ is replaced by $(\mu v^2 /2)(M V^2/2)$ in Eq. (17) of Ref.~\cite{Bertulani:2012sv}.  As a result, they separate the distribution functions of the relative velocity and the CM velocity. Their calculations below Eq. (17) and Eq. (13) which are based upon Eq. (17) are then inconsistent with momentum conservation in Tsallis statistics. \footnote{In addition, Figure 9 in Ref. \cite{Bertulani:2012sv} is difficult to interpret physically. For larger $q$ values, nuclei with low energies are more abundant and those with high energies are less. Since the $\sigma v$ value for $^7$Be($n$,$p$)$^7$Li is larger for smaller CM energy, the thermal reaction rate is expected to be larger for larger $q$. This expectation is opposite to their Fig. 9. In any case, we do not expect a difference by $\lesssim 10$ orders of magnitude in the reaction rate since the $\sigma v$ value is of the same order of magnitude in the wide energy region $E\lesssim 10$ MeV and the nuclear number density should be unchanged (see D. Jang, et al., in preparation). We note that in Ref. \cite{Bertulani:2012sv}, they obtained a similar large effect on the rate of $^3$He($n$,$p$)$^3$H as that of $^7$Be($n$,$p$)$^7$Li (their Sec. 4.1).}  Equations for thermonuclear reaction rates are given in a subsequent paper \cite{Hou:2017uap} [their Eqs. (3) and (4)].  These  are based upon the same formulation  as in Ref.~\cite{Bertulani:2012sv}.  Therefore, for the case of a pure Tsallis distribution the results in \cite{Hou:2017uap} are also inconsistent with momentum conservation. As noted above, although the forms of the distribution function are the same for $\bfv_i$ and $\bfv$ in the MB case, they are different in the case of general statistics.

The thermonuclear reaction rate for Tsallis statistics is then given by
\begin{equation}
  \langle \sigma v \rangle =\int d \bfv f_q^{\rm rel}(\bfv)
  \sigma v,
  \label{eq27}
\end{equation}

When the particles are nonrelativistic, i.e., $x_i =m_i c^2/(kT) \gg1$, the $B_q(x_i)$ factor does not depend on $x_i$. In the present case of the nuclear distribution during the BBN epoch, nuclei are nonrelativistic. In this case the distribution function of $v$ reduces to
\begin{widetext}
\begin{eqnarray}
  f_q^{\rm rel}(\bfv) &=& 
  \frac{B_q^2}{\left( 2 \pi \right)^2}
  \left(x_1 x_2 \right)^{3/2}
  \int_{-1}^1 d \cos\theta \int_0^{V_{\rm max}} V^2 dV
  I_q(V, \cos\theta; m_1, m_2, T, v), \nonumber \\
  \\
  \label{eq28}
  I_q(V, \cos\theta; m_1, m_2, T, v) &=&
  \left\{
  \begin{array}{l}
  \left[1 -\left( q -1 \right) \frac{x_1 v_1^2}{2} \right]^{1/(q-1)}
  \left[1 -\left( q -1 \right) \frac{x_2 v_2^2}{2} \right]^{1/(q-1)} \\
  ~~~~~~~~~~(1 -\left( q -1 \right) \frac{x_1 v_1^2}{2c^2} > 0~{\rm and}~
  1 -\left( q -1 \right) \frac{x_2 v_2^2}{2c^2} >0) \\
  0~~~~~~~~~~({\rm otherwise}) \\
  \end{array}
  \right. \\
  \label{eq29}
  v_1^2 &=& V^2 +2 \frac{m_2}{M} V v \cos\theta +\frac{m_2^2}{M^2} v^2 \\
  \label{eq30}
  v_2^2 &=& V^2 -2 \frac{m_1}{M} V v \cos\theta +\frac{m_1^2}{M^2} v^2,
  \label{eq31}
\end{eqnarray}
\end{widetext}
where
we adopt the natural units of
$k =c =1$.

\subsection{Reduced equations}\label{sec2c}
\subsubsection{Tsallis statistics}\label{sec2c_1}
We defined a thermal velocity $v_{\rm th}$ such that the distribution function for the relative velocity amplitude $f^{\rm rel}(v)$ for MB statistics is maximal at that velocity. The distribution function $f(v)$ satisfies $\int f(v) dv =\int f(\bfv) d\bfv =4\pi \int f(\bfv) v^2 dv$ under the assumption of isotropy\footnote{We note that distribution functions of the velocity $v$ and the velocity vector $\bfv$ should not be confused throughout the formulation.}. The thermal velocity is given by
\begin{equation}
  v_{\rm th} = \sqrt{\mathstrut \frac{2kT}{\mu}}.
  \label{eq32}
\end{equation}
We then normalize all velocity variables in terms of the thermal velocity as follows:
\begin{eqnarray}
  y &=& \frac{V}{v_{\rm th}}
  \label{eq33} \\
  y_{\rm max} &=& \frac{V_{\rm max}}{v_{\rm th}}
  =\left\{
    \begin{array}{ll}
      \frac{ m_1 \sqrt{\mathstrut m_2} +m_2 \sqrt{\mathstrut m_1}}
           {\sqrt{\mathstrut q-1} M^{3/2}} & (\mathrm{for}~q >1) \\
           \infty & (\mathrm{for}~q\le 1) \\
    \end{array} \right.
  \label{eq34} \\
  r&=& \frac{v}{v_{\rm th}} =\sqrt{\mathstrut \frac{E}{kT}}
  \label{eq35} \\
  r_i&=& \frac{v_i}{v_{\rm th}}
  \label{eq36}
\end{eqnarray}
\begin{eqnarray}
  r_1(y, \cos\theta; m_1, m_2, r)^2 &=& y^2 +2 \frac{m_2}{M} y r \cos\theta +\frac{m_2^2}{M^2} r^2 \nonumber \\
  \label{eq37} \\
  r_2(y, \cos\theta; m_1, m_2, r)^2 &=& y^2 -2 \frac{m_1}{M} y r \cos\theta +\frac{m_1^2}{M^2} r^2. \nonumber
  \label{eq38} \\
\end{eqnarray}
Using this transformation, we derive the distribution function for Tsallis statistics, i.e.,
\begin{widetext}
\begin{eqnarray}
  f_q^{\rm rel}(\bfv) &=& 
  \frac{B_q^2}{\left( 2 \pi \right)^2}
  \left( \frac{4m_1 m_2}{\mu^2} \right)^{3/2} v_{\rm th}^{-3}
  \int_{-1}^1 d \cos\theta \int_0^{y_{\rm max}} y^2 dy
  I_q(y, \cos\theta; m_1, m_2, r), \label{eq39} \nonumber \\
  \\
  I_q(y, \cos\theta; m_1, m_2, r) &=&
  \left[1 -\left( q -1 \right) \frac{m_1 r_1^2}{\mu} \right]^{1/(q-1)}
  \left[1 -\left( q -1 \right) \frac{m_2 r_2^2}{\mu} \right]^{1/(q-1)}.
  \label{eq40} 
\end{eqnarray}
\end{widetext}

\subsubsection{the 1 particle Tsallis distribution for the reduced mass}\label{sec2c_2}
Equation (\ref{eq12}) with the mass replaced with $\mu$ is given by
\begin{equation}
  f_q(\bfv) =B_q \frac{1}{\pi^{3/2}} \frac{1}{v_{\rm th}^3}
  \left[1 -\left( q -1 \right) \frac{E}{k T} \right]^{1/(q-1)}.
  \label{eq41} 
\end{equation}
\subsubsection{MB distribution}\label{sec2c_3}
For comparison, the MB distribution function can be written as
\begin{equation}
  f_{\rm MB}(\bfv) =\frac{1}{\pi^{3/2}} \frac{1}{v_{\rm th}^3} 
  \exp\left( -\frac{E}{k T} \right).
  \label{eq42} 
\end{equation}
The quantity corresponding to $I_q$ in the Tsallis statistical case for the MB case is given by
\begin{equation}
  I_{\rm MB}(y; m_1, m_2, r) =
  \exp \left[ - \left( \frac{M}{\mu} y^2 +r^2 \right) \right].
  \label{eq43} 
\end{equation}

We find that the normalized distribution functions $v_{\rm th}^3f^{\rm rel}(\bfv)$ only depend on $E/T$ [and $m_1$ and $m_2$ for the Tsallis case: Eq. (\ref{eq39})] and these shapes do not essentially evolve along with the cosmic expansion.

\subsection{Reverse reaction rates}\label{sec2d}

The detailed balance relations \cite{Blatt1991} between cross sections of forward and reverse reactions for 1(2,$\gamma$)3 and 1(2,3)4 are given by 
\begin{eqnarray}
  \sigma_{3(\gamma,2)1} &=&\frac{g_1 g_2}{(1+\delta_{12}) g_3}
  \left( \frac{\mu_{12} E_{12}}{E_\gamma^2} \right)
    \sigma_{1(2,\gamma)3}
    \label{eq44} \\
    \sigma_{4(3,2)1} &=&\frac{(1+\delta_{34})g_1 g_2 m_1 m_2 E_{12}}
          {(1+\delta_{12}) g_3 g_4 m_3 m_4 E_{34}}
          \sigma_{1(2,3)4},
          \label{eq45}
\end{eqnarray}
where
the $g_i$ are the statistical weights of the respective nuclear species $i$, while 
$\mu_{ij}$ and $E_{ij}$ are respectively the reduced mass and the CM energy of the $i$+$j$ system.

\subsubsection{Photodisintegration reactions}\label{sec2d_1}
Under the assumption that nuclei are nonrelativistic and that photons have a Planckian energy distribution, the photodisintegration rates do not depend on the nuclear distribution function. The photodisintegration rate is then given by
\begin{eqnarray}
  \langle \sigma c \rangle &=&\int d E_\gamma f(E_\gamma)
  \sigma(E_\gamma) c \\
  \label{eq46} 
  f(E_\gamma) &=& \frac{E_\gamma^2 /[ \exp(E_\gamma/kT) -1]}
  {\int dE_\gamma E_\gamma^2/[ \exp(E_\gamma/kT) -1]} \nonumber \\
  &=& \frac{E_\gamma^2 /[ \exp(E_\gamma/kT) -1]}
  {2 \zeta(3) (kT)^3},
  \label{eq47} 
\end{eqnarray}
where
  $E_\gamma$ is the photon energy,
  $f(E_\gamma)$ is the distribution function of the photon energy,
  $\sigma(E_\gamma)$ is the photodisintegration cross section,
$c$ is the light speed, and
$\zeta(3)=1.2021$ is the Riemann zeta function of 3. As far as the photon distribution follows the Planck distribution, the photodisintegration rate is the same as that of the SBBN \footnote{Effects of non-Planck distribution of background radiation will be discussed anywhere else (D. Jang, et al. in preparation).}. Therefore, we adopt the standard rates.

\subsubsection{Two-nuclear reactions}\label{sec2d_2}
The thermal reaction rate for the reverse reaction of the type 4(3,2)1 is
\begin{equation}
  \langle \sigma v \rangle_{34} =\int d \bfv_3 f(\bfv_3) \int d \bfv_4 f(\bfv_4)
  \sigma_{4(3,2)1}(E_{34}) v_{34},
  \label{eq48} 
\end{equation}
where
the subscript 34 indicates physical quantities of the 3+4 system, and
the subscripts 3 and 4 indicate physical quantities of particles 3 and 4, respectively. When the distribution functions for all nuclei are the Tsallis distribution, this rate is reduced with Eqs. (\ref{eq39}) and (\ref{eq40}) to
\begin{equation}
  \langle \sigma v \rangle_{34} =\int d \bfv_{34} f_q^{\rm rel}(\bfv_{34}) 
  \sigma_{4(3,2)1}(E_{34}) v_{34}.
  \label{eq49} 
\end{equation}
We calculate reverse reaction rates using this equation and the detailed balance relation [Eq. (\ref{eq45})].

\section{Results}\label{sec3}

\subsection{Relative velocity distribution}\label{sec3a}

\begin{figure}[tbp]
\begin{center}
\includegraphics[width=8.0cm,clip]{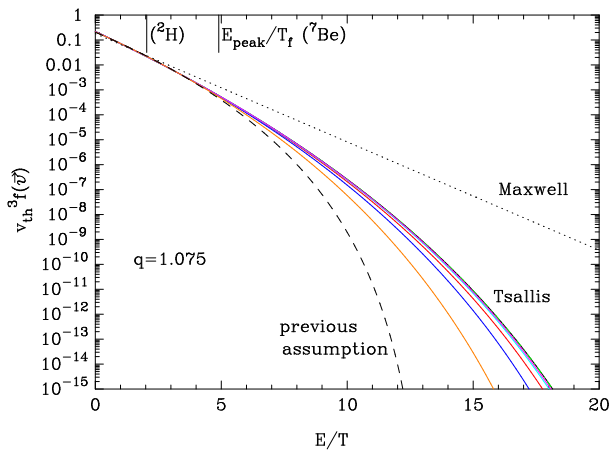}
\caption{Normalized distribution function for the CM kinetic energy $E$ for $q =1.075$. Solid and dash-dotted lines are for Tsallis statistics, i.e., $(A_1, A_2) =(1,1)$, $(2,2)$ (dash-dotted line), $(4,3)$, $(3,2)$, $(2,1)$, $(3,1)$, and $(7,1)$ from the top to the bottom, respectively. The dashed line corresponds to the erroneous distribution taken from the one-particle Tsallis distribution. The dotted line is the MB distribution. Upper vertical lines show the ratios $E_{\rm peak}/T_{\rm f}$ for the freeze-out of $^2$H and $^7$Be nuclides (see Table \ref{tab1}).
\label{fig1}}
\end{center}
\end{figure}

Figure \ref{fig1} shows normalized distribution functions for the CM kinetic energy $E$. The Tsallis parameter is set to $q =1.075$ \cite{Hou:2017uap}. That value has been suggested \cite{Hou:2017uap} as the value for which the Li problem is solved. Solid and dash-dotted lines are functions for the Tsallis statistics [Eqs. (\ref{eq39}) and (\ref{eq40})] for sets of nuclear mass numbers of $(A_1, A_2) =(1,1)$, $(2,2)$ (dash-dotted line), $(4,3)$, $(3,2)$, $(2,1)$, $(3,1)$, and $(7,1)$ from the top to the bottom, respectively. The dashed line corresponds to the previously assumed distribution, i.e., the one-particle Tsallis distribution [Eq. (\ref{eq41})]. The dotted line is the MB distribution [Eq. (\ref{eq42})]. The MB distribution $v_{\rm th}^3 f^{\rm rel}_{\rm MB}(\bfv)$ depends only on $E/T$. The previously assumed distribution $v_{\rm th}^3 f_q(\bfv)$ also depends only on $E/T$ for a fixed $q$. The distribution function for the Tsallis statistics $v_{\rm th}^3 f_q^{\rm rel}(\bfv)$ depends on the nuclear masses as well as $E/T$ for a fixed $q$. This is one of the important differences from MB statistics.


In addition, the Tsallis distribution functions for the relative velocity have extended high energy tails compared with the previously assumed function. The maximum velocity [Eq. (\ref{eq22})] is always larger than that of the one particle Tsallis distribution for the same reduced mass. Therefore, the extended high energy tails are realized in the exact distribution function.

  We note that the assumption of the the same Tsallis distribution for the relative velocity [Eq. (\ref{eq41})] independent of masses of reacting nuclei \cite{Bertulani:2012sv,Hou:2017uap} induces an inconsistency if $q \ne 1$ and therefore unphysical. As shown above, the relative velocity distribution is derived from the velocity distribution functions of nuclei, and its relation to the nuclear distribution functions depends on the nuclear masses. If one supposes that the relative velocity distribution, which is in fact a physical quantity derived from distribution functions of respective nuclear velocities,  generally follows the Tsallis form, solutions for distribution functions of respective nuclei cannot be found except for the trivial case of $q=1$.

\subsection{Thermonuclear reaction rates}\label{sec3b}

Table \ref{tab1} shows the eleven important reactions of BBN \cite{Kawano1992,Smith:1992yy} (the first column), and references to the available cross section data that we adopted in this study (the second column).

We calculated the freeze-out temperature $T_{\rm f}$ [$T_{9{\rm f}}=T_{\rm f}/(10^9~{\rm K})$] for one chosen nuclide $i$ participating in one reaction $a$ which satisfies the freezeout condition that its abundance rate of change equals the cosmic expansion rate $H(T)$, i.e.,
\begin{equation}
  H(T_{\rm f}) =\frac{\left| (dn_i/dt)_a \right|}{n_i}
  =\frac{n_k n_l \langle \sigma v (T_{\rm f}))\rangle_{kl}}{n_i},
  \label{eq50} 
\end{equation}
where
$n_j$ is the number density of nuclide $j$, while
$k$ and $l$ are nuclides in the initial state of the reaction $a$,
and
$\langle \sigma v (T))\rangle_{kl}$ is the average reaction rate as a function of $T$.
The third and fourth columns in Table \ref{tab1} show the nuclide whose abundance freezes out and its freeze-out temperature, respectively. The fifth column shows the ratio of the peak energy to $T_{\rm f}$, where the peak energy has the largest contribution to the integrand in deriving the average reaction rate. We checked that the change in the peak energy is small if the Tsallis $q$ value is not changed significantly, i.e., $|q-1| \lesssim 0.1$.

\begin{table}
  \caption{\label{tab1} Important BBN reactions}
  \begin{ruledtabular}
    \begin{tabular}{lllll}
      reaction & reference & nuclide & $T_{9{\rm f}}$ & $\frac{E_{\rm peak}}{T_{{\rm f}}}$ \\
      \hline
$^1$H($n$,$\gamma$)$^2$H        & \cite{Ando:2005cz} (Fig. 3)                                     & $^2$H  & 0.53 & 0.68 \\
$^3$He($n$,$p$)$^3$H            & \cite{Descouvemont:2004cw}                                      & $^3$He & 0.54 & 0.78 \\
$^7$Be($n$,$p$)$^7$Li           & \cite{Descouvemont:2004cw}                                      & $^7$Be & 0.44 & 0.66 \\
$^2$H($p$,$\gamma$)$^3$He       & \cite{Coc:2015bhi} (\cite{Descouvemont:2004cw} for $E$ > 2 MeV)   & $^2$H  & 0.74 & 2.1 \\
$^7$Li($p$,$\alpha$)$^4$He      & \cite{Descouvemont:2004cw}                                      & $^7$Li & 0.18 & 4.3 \\
$^2$H($d$,$p$)$^3$H             & \cite{Coc:2015bhi} (\cite{Descouvemont:2004cw} for $E$ > 0.6 MeV) & $^2$H  & 0.55 & 2.0 \\
$^2$H($d$,$n$)$^3$He            & \cite{Coc:2015bhi} (\cite{Descouvemont:2004cw} for $E$ > 0.6 MeV) & $^2$H  & 0.53 & 2.1 \\
$^3$H($d$,$n$)$^4$He            & \cite{Descouvemont:2004cw}                                      & $^3$H  & 0.12 & 3.2 \\
$^3$He($d$,$p$)$^4$He           & \cite{Descouvemont:2004cw}                                      & $^3$He & 0.40 & 3.5 \\
$^3$H($\alpha$,$\gamma$)$^7$Li  & \cite{Descouvemont:2004cw}                                      & $^7$Li & 0.21 & 4.1 \\
$^3$He($\alpha$,$\gamma$)$^7$Be & \cite{Descouvemont:2004cw}                                      & $^7$Be & 0.41 & 4.9 \\ 
    \end{tabular}
  \end{ruledtabular}
\end{table}

\begin{figure}[tbp]
\begin{center}
\includegraphics[width=8.0cm,clip]{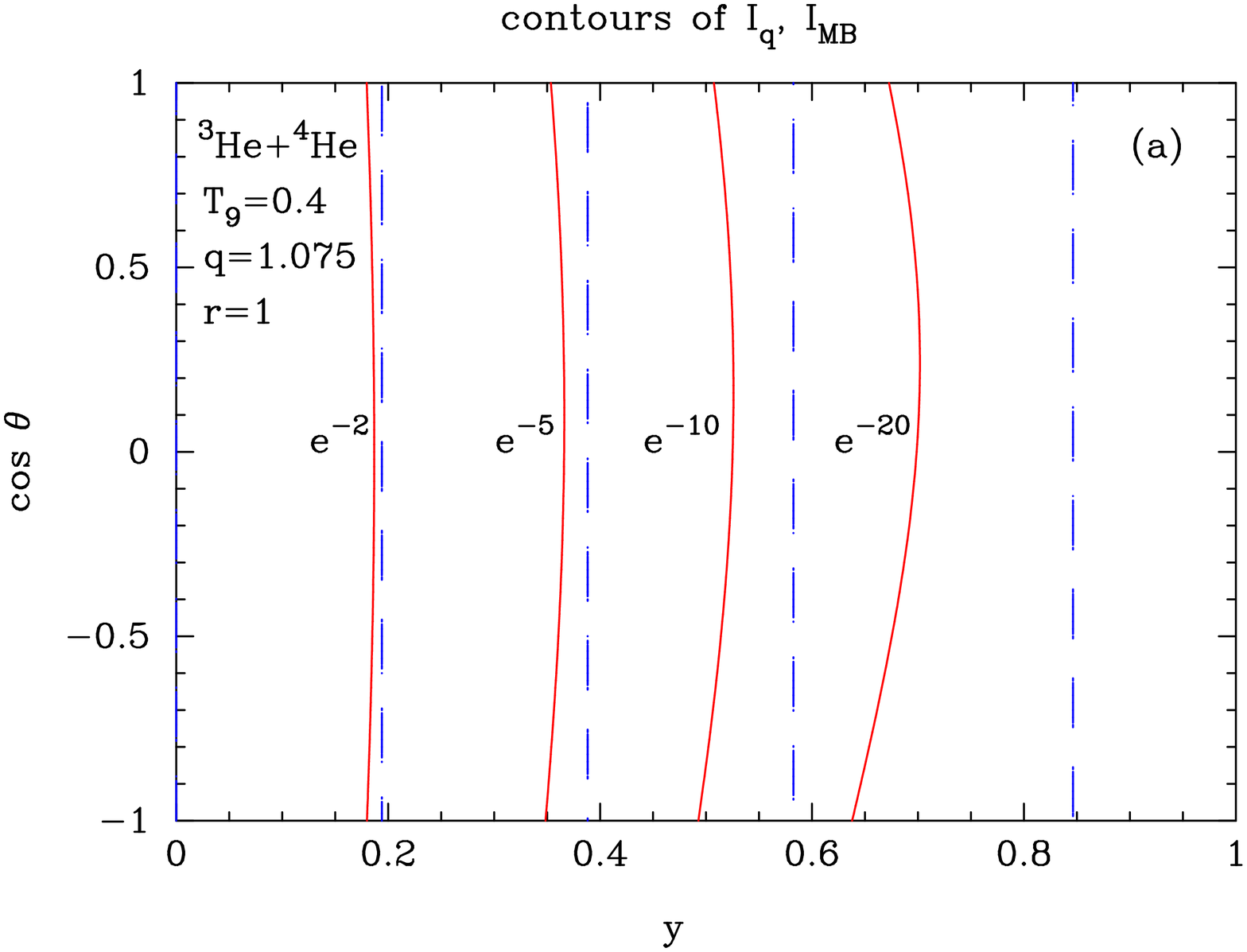}
\includegraphics[width=8.0cm,clip]{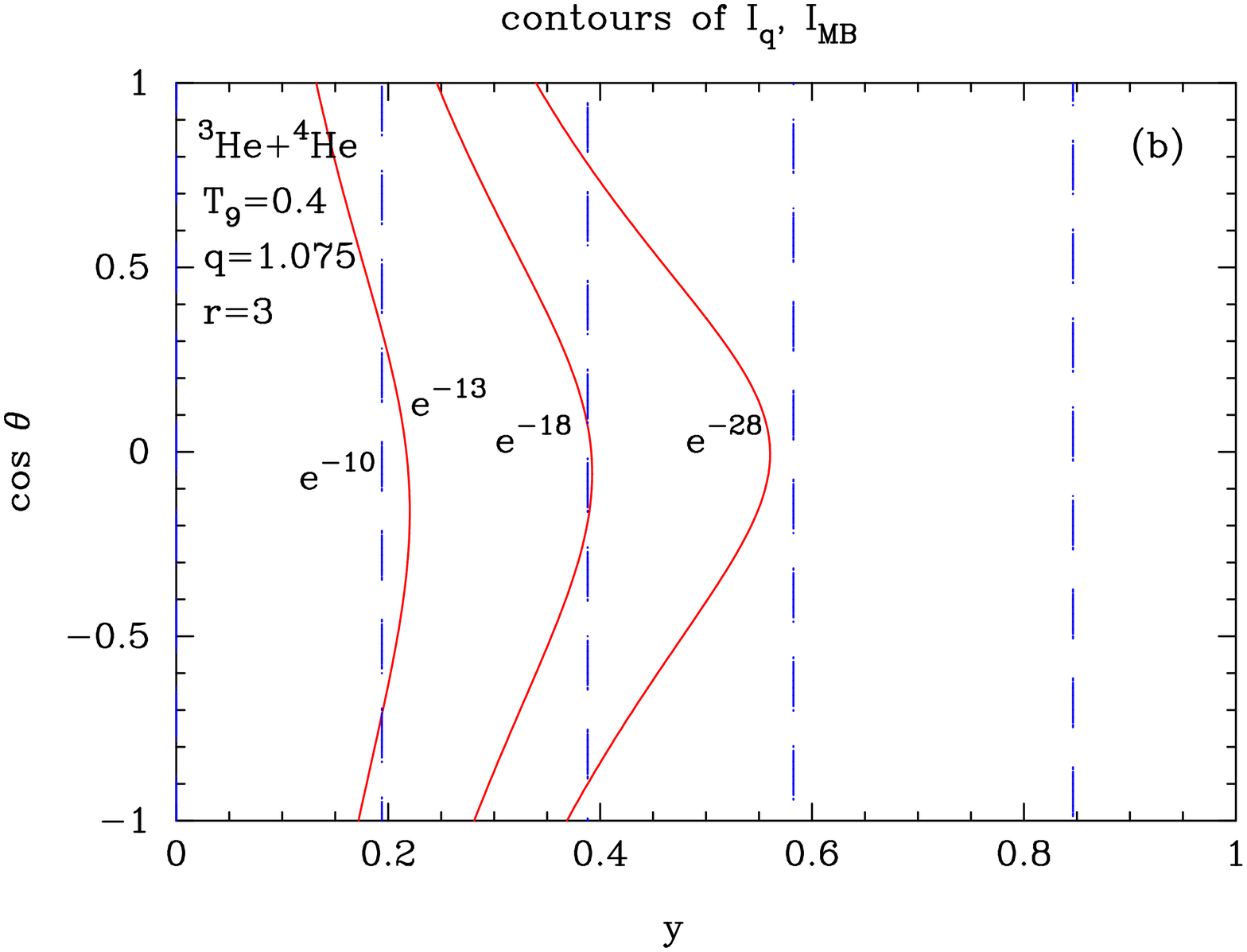}
\caption{Contours of the functions $I_{q=1.075}$ and $I_{\rm MB}$ in the ($y$,$\cos\theta$) parameter plane for the $^3$He+$^4$He system at $T_9 =0.4$. Panel (a) corresponds to the energy $r=1$. The solid and dashed lines from left to right show contours of $e^{-2}$, $e^{-5}$, ... $e^{-20}$, as labeled. Panel (b) is for $r=3$. The solid lines are for $I_{q=1.075}$ values of $e^{-13}$, $e^{-18}$, and $e^{-28}$, respectively, while the dashed lines are for $I_{\rm MB}=e^{-10}$, $e^{-13}$, ..., $e^{-28}$ from the left to right, respectively.
  \label{fig2}}
\end{center}
\end{figure}

  Figure \ref{fig2} shows contours of the functions $I_{q=1.075}$ [Eq. (\ref{eq40})] (solid lines) and $I_{\rm MB}$ [Eq. (\ref{eq43})] (dashed lines) in the ($y$,$\cos\theta$) plane for the $^3$He+$^4$He system at $T_9 =0.4$ for the energies of $r=1$ (a) and 3 (b), respectively. From this figure, a difference in the contributions of the parameter regions to the reaction rate between the Tsallis and MB cases is apparent. We considered the $^7$Be production reaction and its freeze-out temperature $T_{9{\rm f}} =0.4$ for this figure. In panel (a), it is seen that $I_q$ values are hindered compared to the MB case. In addition, the $I_q$ value significantly depends on the angle $\theta$ between \bfV and \bfv, which is different from the function $I_{\rm MB}$. Near parallel or anti-parallel scatterings are hindered as seen from the curved contours of $I_q$. In the regions of $\cos\theta \gtrsim -1$ and $\cos\theta \lesssim 1$, $r_1^2$ or $r_2^2$ becomes maximally large leading to small values of $I_q$ [Eq. (\ref{eq40})]. For intermediate $\cos\theta$ values, both of $r_1^2$ and $r_2^2$ values are intermediate, and the hindrance of $I_q$ is minimal.
  In panel (b), vertical dashed lines correspond to $I_{\rm MB}=e^{-10}$, $e^{-13}$, ..., $e^{-28}$, from the left to right, respectively, that is a hindrance factor depending on $y$, exactly the same as those in panel (a). The solid lines show the contours of $I_{q=1.075}=e^{-13}$, $e^{-18}$, and $e^{-28}$, respectively. Curvatures are larger than in panel (a), which indicates that the contribution from the intermediate $\cos\theta$ satisfying $m_1 r_1^2 =m_2 r_2^2$ is exclusively important.


\begin{figure}[tbp]
\begin{center}
\includegraphics[width=8.0cm,clip]{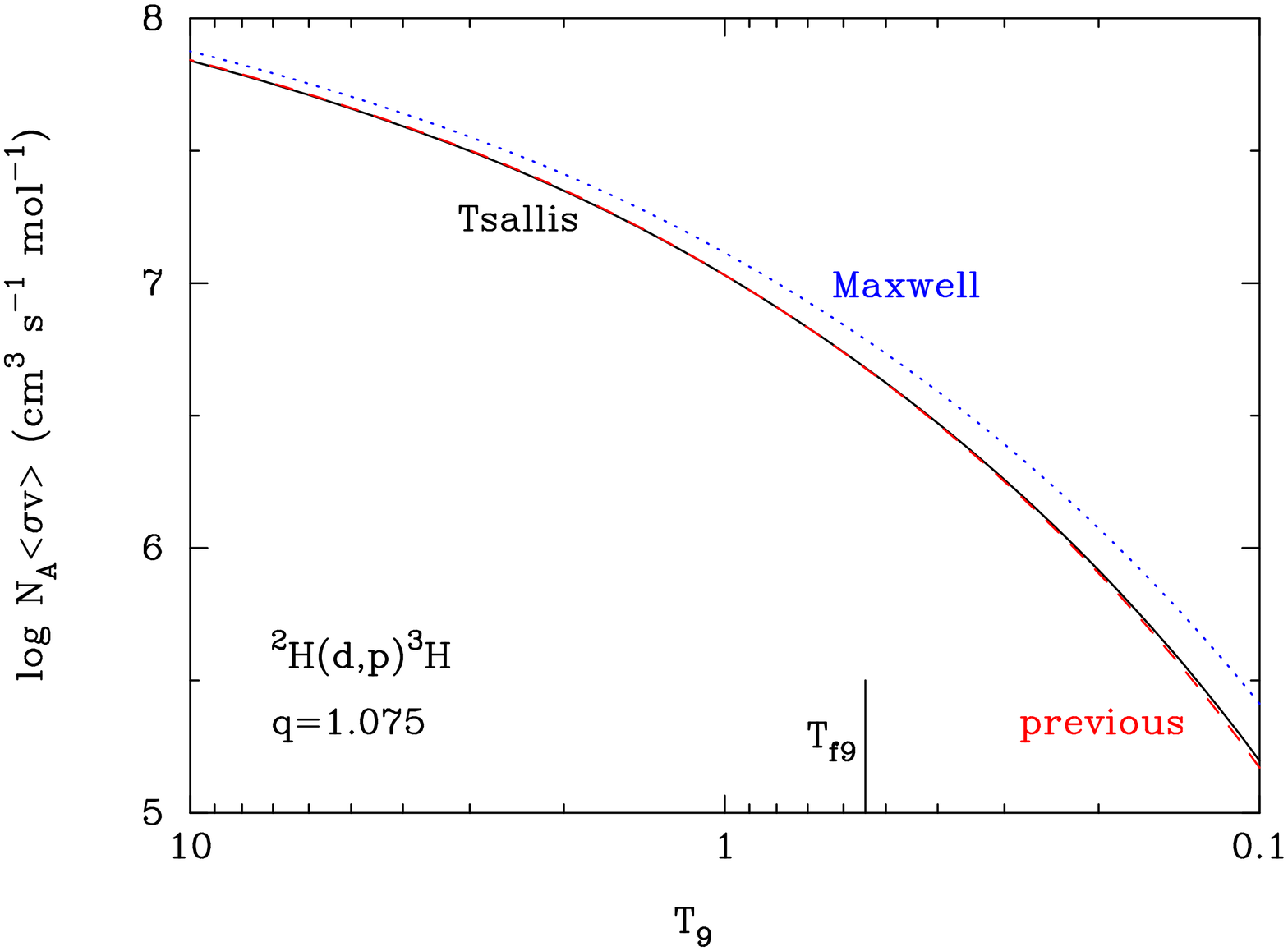}
\includegraphics[width=8.0cm,clip]{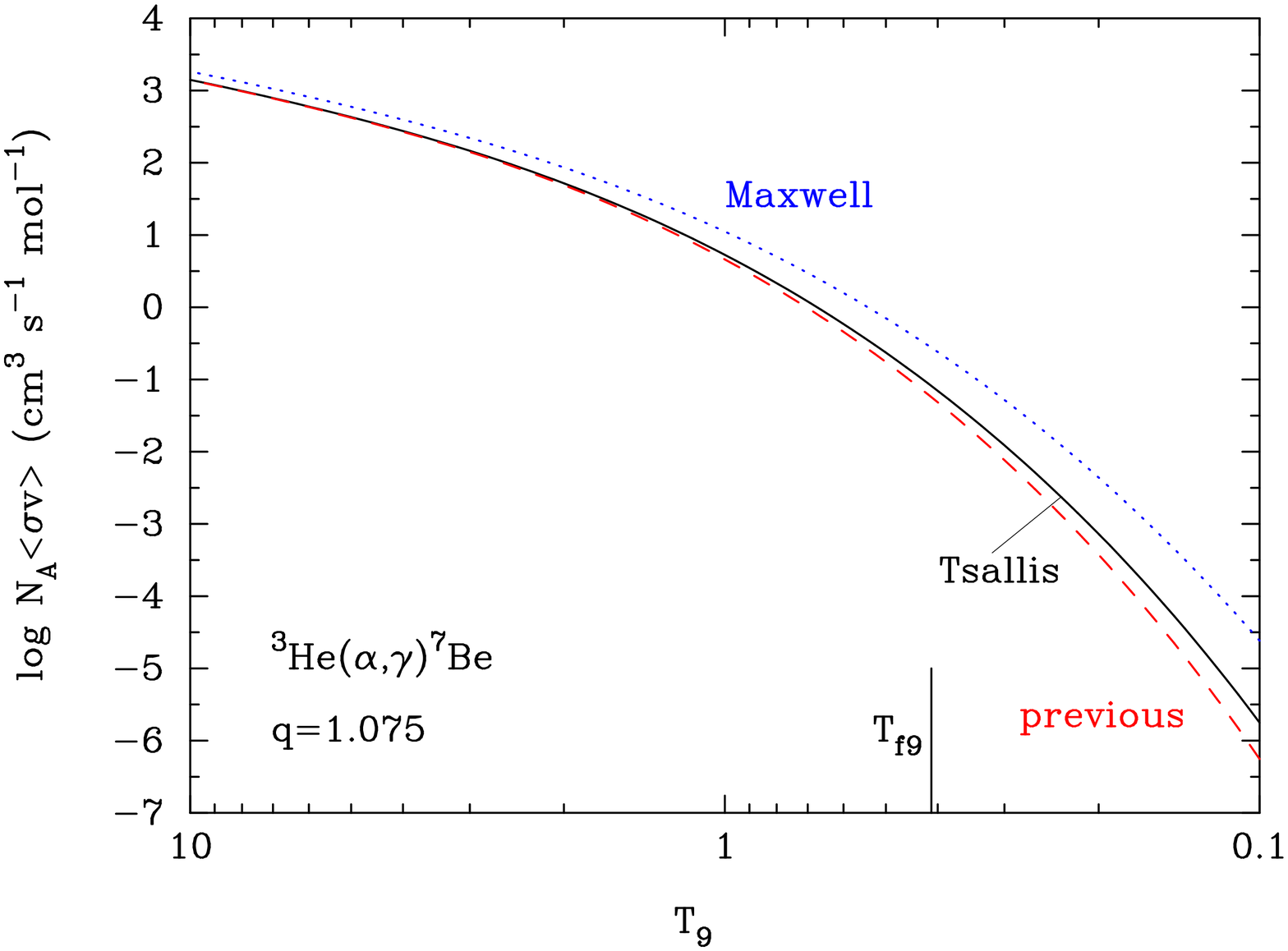}
\caption{The average rates for the reactions $^2$H($d$,$p$)$^3$H (upper panel) and $^3$He($\alpha$,$\gamma$)$^7$Be (lower panel) as a function of $T_9$. The Tsallis parameter is set to $q =1.075$. The solid line is for the Tsallis statistics, while the dashed line is erroneously based on the one-particle distribution function. The dotted line corresponds to the MB distribution. The lower vertical lines show the freeze-out temperatures of $^2$H (upper panel) and $^7$Be (lower panel), respectively (see Table \ref{tab1}).
  \label{fig3}}
\end{center}
\end{figure}

Figure \ref{fig3} shows the average rates for the reactions $^2$H($d$,$p$)$^3$H (upper panel) and $^3$He($\alpha$,$\gamma$)$^7$Be (lower panel) as a function of temperature $T_9 \equiv T/(10^9~{\rm K})$. The former reaction is one of two most important reactions for D destruction. We note that the reaction rate of the other reaction, i.e., $^2$H($d$,$n$)$^3$He is very similar to that of $^2$H($d$,$p$)$^3$H. The $^3$He($\alpha$,$\gamma$)$^7$Be reaction is the most important $^7$Be production reaction. Therefore, the two reactions in Fig. \ref{fig3} determine the final freeze-out abundances of D and $^7$Be, respectively. The solid line is for the Tsallis statistics, while the dashed line is based upon the one-particle distribution function as previously assumed. The dotted line corresponds to the MB distribution.


For this specific $q$ value and the temperature range of $T_9=[10,0.1]$, the $^2$H($d$,$p$)$^3$H reaction rate for the Tsallis statistics is smaller than that of the MB statistics. Since the distribution function at high energies is smaller than that of the MB function, the reaction rate is hindered by the Coulomb penetration factor. The difference between cases of the Tsallis statistics and the previously assumed function is small.

The reaction rate of $^3$He($\alpha$,$\gamma$)$^7$Be for the Tsallis statistics is also smaller than that of the MB statistics because of the more effective Coulomb suppression factor. We find a significant difference between the reaction rates of the Tsallis statistics and the previously assumed case. For a fixed CM energy, the Coulomb penetration factor is suppressed more than that of the reaction $^2$H($d$,$p$)$^3$H because of the larger atomic numbers and reduced mass. Therefore, the thermal reaction rate is contributed from higher energies, where the difference in the distribution function $f^{\rm rel}(\bfv)$ between the Tsallis and the previous case is largest (see Fig. \ref{fig1}). The averaged reaction rates of $^3$He($\alpha$,$\gamma$)$^7$Be then differ more than those of $^2$H($d$,$p$)$^3$H.

\subsection{BBN calculation}\label{sec3c}

We adopt the SBBN code as described in Refs.~\cite{Kawano1992,Smith:1992yy} and have updated reaction rates of nuclei with mass numbers $\le  10$ using the JINA REACLIB Database \cite{Cyburt2010} (updated to December, 2014). The neutron lifetime is the central value of the Particle Data Group, $880.2 \pm 1.0$~s~\cite{2016ChPhC..40j0001P}.  The baryon-to-photon ratio is taken to be $(6.094 \pm 0.063) \times 10^{-10}$ calculated using a conversion \cite{Ishida:2014wqa} of the baryon density in the standard $\Lambda$CDM model (TT+lowP+lensing) determined from the Planck observation of the cosmic microwave background, $\Omega_\mathrm{m} h^2 =0.02226 \pm 0.00023$ \cite{Ade:2015xua}. For eleven important reactions of BBN, the reaction rates are calculated for the different distribution cases. Two-body reverse reaction rates are calculated with the detailed balance relation using Eqs. (\ref{eq44})--(\ref{eq47}) and (\ref{eq49}). Since the effect of the Planckian distribution is small \cite{Mathews:2010aa}, we make the usual approximation of replacing the Planck distribution with an exponential.

\begin{figure}[tbp]
\begin{center}
\includegraphics[width=8.0cm,clip]{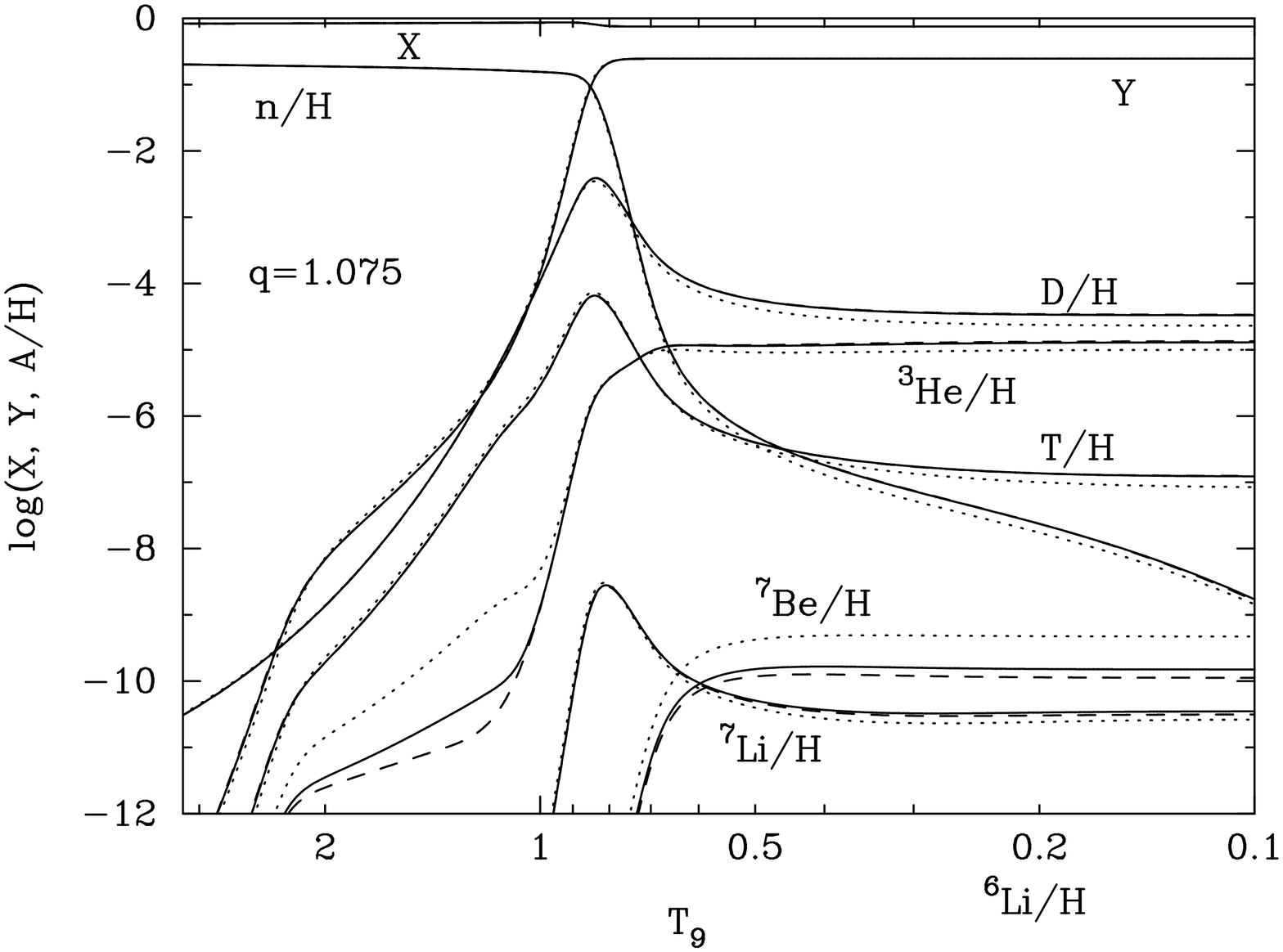}
\includegraphics[width=8.0cm,clip]{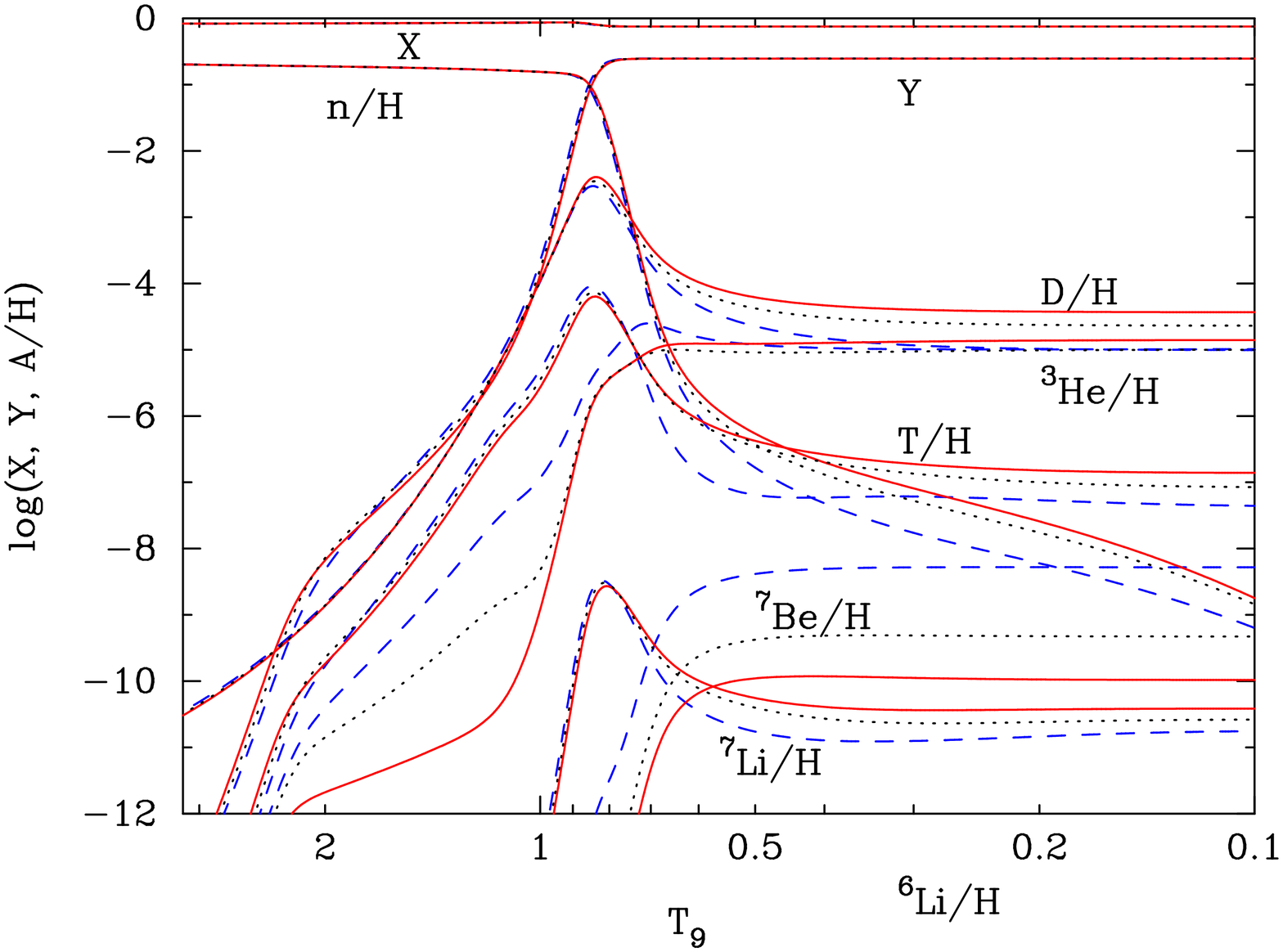}
\caption{Evolution of nuclear abundances as a function of $T_9$. $X$ and $Y$ are mass fractions of $^1$H and $^4$He in total baryon matter, respectively. Other nuclear abundances are shown by the number ratios to $^1$H, i.e., $A$/H. In the upper panel, the solid lines are for the Tsallis statistics, while the dashed lines are for the previously assumed relative velocity distribution function. The dotted lines are results for the MB statistics. The Tsallis parameter is set to $q =1.075$. The lower panel shows abundances for Tsallis statistics with $q =0.9$ (dashed lines), 1 (dotted), and 1.1 (solid), respectively.
  \label{fig4}}
\end{center}
\end{figure}

Figure \ref{fig4} shows the evolution of nuclear abundances as a function of $T_9$. $X$ and $Y$ are mass fractions of $^1$H and $^4$He in total baryon matter, respectively. For other nuclear abundances, the number ratios to $^1$H, i.e., $A$/H are shown. In the upper panel, the solid lines are for the Tsallis statistics, while the dashed lines are for the previously assumed relative velocity distribution function. The dotted lines are results for the MB statistics, i.e., SBBN. The Tsallis parameter is set to $q =1.075$.

For $q>1$, the high energy region of the velocity distribution functions of nuclei is suppressed. Therefore, rates of charged-particle reactions are significantly reduced since the cross sections are larger at high energies. For $q<1$, the opposite situation is realized. On the other hand, rates of neutron reactions are unaffected because there is no Coulomb penetration factor.

In the upper panel of Fig. \ref{fig4}, since the reaction rate of deuteron destruction via $^2$H($d$,$p$)$^3$H and $^2$H($d$,$n$)$^3$He is smaller than in SBBN (Fig. \ref{fig3}, upper panel), the freeze-out D abundance is larger \cite{Bertulani:2012sv}. Since $^3$He and $^3$H are produced via the same reactions, the higher D abundance results in higher production rates of $^3$He and $^3$H. Therefore, the final abundances of $^3$He and $^3$H are higher than in SBBN. The neutron abundance is slightly larger at late times for which $T_9 \lesssim 0.7$. This is because of the larger D abundance. The neutron abundance is determined from the forward and reverse reactions of $^1$H($n$,$\gamma$)$^2$H. The forward rate is slightly larger and the reverse rate is the same as that of the SBBN. The $^1$H abundance is almost the same in the Tsallis case and SBBN. Therefore, after the D destruction freezes out at a higher level, the $n$ abundance is kept higher via the photodisintegration $^2$H($\gamma$,$n$)$^1$H. The final $^7$Be abundance is smaller than in SBBN because of the smaller reaction rate for $^3$He($\alpha$,$\gamma$)$^7$Be (Fig. \ref{fig3}, lower panel). Since the reaction rate is underestimated in the previous calculations \cite{Bertulani:2012sv,Hou:2017uap}, our $^7$Be abundance is larger than the previous estimate. The $^7$Li destruction rate via $^7$Li($p$,$\alpha$)$^4$He is also smaller than in SBBN. Therefore, the freeze-out $^7$Li abundance is larger. 

The lower panel shows abundances for Tsallis statistics with $q =0.9$ (dashed line), 1 (dotted), and 1.1 (solid), respectively. When the $q$-value increases, rates of charged-particle reactions are smaller. For the same reason explained for the upper panel, the abundances of D, $^3$H, $^3$He, $n$, and $^7$Li increase while the $^7$Be abundance decreases.


Calculated BBN results are compared to observational constraints on light element abundances.
Constraints on the primordial $^4$He abundance come from observations of metal-poor extragalactic H II regions. We use two different determinations of $Y_{\rm p}=0.2446 \pm 0.0029$ \cite{Peimbert:2016bdg} and $Y_{\rm p} =0.2551\pm 0.0022$~\cite{Izotov:2014fga}.
The primordial D abundance is constrained with observations of metal-poor Lyman-$\alpha$ absorption systems towards quasi-stellar objects. We use the weighted mean value of D/H$=(2.527 \pm 0.030) \times 10^{-5}$~\cite{Cooke:2017cwo}.
$^3$He abundances in Galactic H II regions are determined using the $8.665$~GHz
hyperfine transition of $^3$He$^+$ ion. The primordial $^3$He abundance can evolve during Galactic chemical evolution. However, the net effect of Galactic chemical evolution is uncertain since stars can both destroy and synthesize $^3$He. Moreover, it is not expected that the $^3$He abundance has decreased significantly over Galactic history as this would require that a large fraction of Galactic baryonic material has been absorbed in stars that destroy $^3$He, while the present interstellar deuterium abundance limits the amount of astration to not more than about a factor of two. We then only adopt the $2\sigma$ upper limit from the abundance $^3$He/H=$(1.9\pm 0.6)\times 10^{-5}$~\cite{Bania:2002yj} in Galactic H II regions. 
We also use the abundance $\log(^7$Li/H)$=-12+(2.199\pm 0.086)$ derived by observations of Galactic metal-poor stars using a 3D nonlocal thermal equilibrium model~\cite{Sbordone2010}.

\begin{figure}[tbp]
\begin{center}
\includegraphics[width=8.0cm,clip]{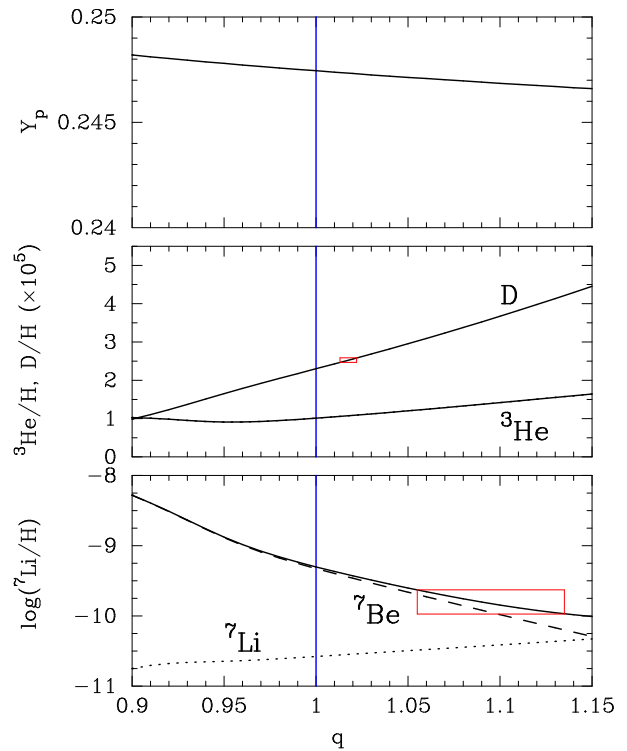}
\caption{Light element abundances versus the Tsallis parameter $q$. Boxes show the 2 $\sigma$ observational limits on D/H and $^7$Li/H. In the panel for $^7$Li/H, dashed and dotted lines show abundances of $^7$Be and $^7$Li, respectively, immediately after the BBN. The vertical line is at $q=1$, i.e., the SBBN case.
  \label{fig5}}
\end{center}
\end{figure}

Figure \ref{fig5} shows calculated light element abundances as a function of the parameter $q$. Boxes show the 2 $\sigma$ observational limits on D/H and $^7$Li/H. The dashed and dotted lines show abundances of $^7$Be and $^7$Li, respectively, immediately after BBN. Long after BBN, $^7$Be nuclei electron capture to produce $^7$Li nuclei. Therefore, the sum of $^7$Be and $^7$Li abundances becomes the primordial Li abundance. The $^3$He abundance is predominantly contributed by $^3$He, plus a small abundance of $^3$H produced during BBN (see Fig. \ref{fig4}) has been added. The vertical line is at $q=1$ and corresponding to the SBBN case. The plotted range is allowed by the 2 $\sigma$ limit on the $^4$He abundance of Ref. \cite{Peimbert:2016bdg} and excluded by that of Ref. \cite{Izotov:2014fga}. Also this region is allowed by the 2 $\sigma$ upper limits on $^3$He/H~\cite{Bania:2002yj}.


The reasons for the abundance changes of D, $^3$He, $^7$Be, and $^7$Li have been explained above. The one percent level of change for the final $^4$He mass fraction $Y_{\rm p}$ is caused by different neutron abundances during the $^4$He synthesis. For larger $q$, the D destruction rate is smaller and the D abundance is larger. As a result of the balance of forward and reverse reactions of $^1$H($n$,$\gamma$)$^2$H (see above), the $n$ abundance is kept higher and more neutrons are lost by $\beta$-decay before $^4$He synthesis is completed. The final $^4$He abundance is therefore smaller.

It is seen that the D abundance increases and the $^7$Li abundance decreases with increasing $q$ value. At $q \approx 1.01$--1.02, the theoretical result for the  D abundance is consistent with the observation. On the other hand, for $q \gtrsim 1.055$, the $^7$Li abundance agrees with the observation. However, in this parameter region, the D abundance is enhanced to above D/H$=3\times 10^{-5}$, which requires an additional mechanism for later D destruction. Because of their fragility, deuterons can be destroyed easily if there is a source of non-thermal photons in the early universe (e.g., \cite{Lindley1979MNRAS.188P..15L,Kusakabe:2013sna}). Then, the D destruction by non-thermal photons can reproduce the primordial elemental abundances consistent with observations of all light nuclei.  This can happen for example,  in a model including  photon cooling by an axion condensate \cite{Kusakabe:2012ds}.

Unless a later D destruction mechanism is induced, the D enhancement is very problematic. Therefore, the observed D abundance places an upper limit on $q$. We find that in the range of $q \approx 1.01$--1.02 where the observed D abundance is reproduced, the $^7$Li abundance is smaller than in the SBBN by $\sim$30--60 \%. This level of Li reduction is significant but not enough. On the other hand, there are other astrophysical processes which can further reduce the Li abundance, i.e., a chemical separation of $^7$Li$^+$ ions during structure formation \cite{Kusakabe:2014dta} or a depletion on stellar surfaces \cite{Richard:2004pj,Korn:2007cx,Lind:2009ta}.

\section{Summary}\label{sec4}

We have reformulated the thermal rate of two-body reactions of a gas of nonrelativistic uncorrelated particles with general velocity distribution functions. Taking the Tsallis distribution as an example of a non-MB distribution, we derived the distribution function of the relative velocity, i.e., $f_q^{\rm rel}(\bfv)$. It was found that in general the distribution function $f^{\rm rel}(\bfv)$ contains a complicated integration over the CM velocity $\bfV$. By defining a normalized distribution function $v_{\rm th}^3 f^{\rm rel}(\bfv)$, the MB distribution can be expressed in terms of the ratio of the CM energy to the temperature, i.e., $E/T$. However, we found that the normalized distribution function for the Tsallis statistics has additional dependences on the nuclear masses (Sec. \ref{sec2}). 

We showed differences in the relative velocity distribution function between the Tsallis and MB statistics (Sec. \ref{sec3a}). Using calculated distribution functions, reaction rates that are important for BBN were derived (Sec. \ref{sec3b}). Finally, we performed the BBN nuclear reaction network calculation, and analyzed effects of changing the Tsallis $q$ parameter. An increase of $q$ results in softer nuclear spectra, and an upper cutoff of the CM energy appears for $q>1$. We observed that the increase of $q$ reduces rates of reactions between charged particles, and explained reasons that abundances of D, $^3$H, $^3$He, $n$, and $^7$Li increase while the $^7$Be abundance decreases. Predicted abundances as a function of the parameter $q$ were calculated. We found the following points: (1) A slight deviation from the MB statistics, i.e., $q \approx 1.01$--1.02 can lead to D abundances consistent with observations, which are larger than the SBBN prediction; (2) The D observation provides the most stringent constraint on the $q$ parameter; (3) In that $q$ region, the primordial Li abundance is reduced from the SBBN value by $\sim$30--60 \% (Sec. \ref{sec3c}).

\begin{acknowledgments}
  MK acknowledges support by the visiting scholar program of NAOJ during his stay there.
  This work was supported by Grants-in-Aid for Scientific Research of JSPS (15H03665, 17K05459).
Work of GJM supported in part by DOE nuclear theory grant DE-FG02-95-ER40934 and in part by the visitor program at NAOJ.
\end{acknowledgments}



\end{document}